\documentclass[aps,prb,preprint,floatfix]{revtex4-1}

\usepackage{amsmath} 
\usepackage{graphicx} 
\usepackage{epsfig}

\begin{document}

\title{The exit velocity of a compressed air cannon}

\author{Z. J. Rohrbach}
\author{T. R. Buresh}
\author{M. J. Madsen}
\affiliation{Department of Physics, Wabash College, Crawfordsville, IN  47933}

\date{\today}

\begin{abstract}
The use of compressed air cannons in an undergraduate lab provides a way to illustrate the cooperation of diverse physics concepts, such as conservation of momentum, the work-kinetic energy theorem, expansion of gas, air drag, and elementary Newtonian mechanics.  However, recent proposals have disagreed as to whether the expansion of the gas in the cannon should be modeled as an adiabatic or an isothermal process.  We built an air cannon that utilized a diaphragm valve to release our pressurized gas and found that neither model accurately predicted the exit velocity of our projectile.  We present a new model, based on the flow of air through the valve, that is in much better agreement with our data.  
\end{abstract}

\maketitle

Although the description of the internal dynamics of a firearm is a complicated task, recent proposals have focused on modeling the dynamics of a simplified cannon which uses the expansion of compressed gas to accelerate a projectile.  However, these proposals disagree about whether the the gas expansion should be described as an adiabatic\cite{m} or an isothermal\cite{d} process.  Thus, these models disagree in their predictions of the exit velocity of a projectile as a function of the initial gas pressure.  Because we are interested in developing an undergraduate physics lab that would use a compressed gas cannon to illustrate conservation of momentum\cite{t} and the work-kinetic energy theorem\cite{ts}, we wanted to have an accurate model that predicts the internal dynamics of the cannon.  We also note that this work could be extended to investigate projectile motion and air drag using elementary Newtonian kinematics.\cite{k}  

We constructed a compressed air cannon to measure the exit velocity of the projectile as a function of the initial reservoir pressure.  We report that neither the adiabatic nor the isothermal model accurately predicts the exit velocity of a projectile from our cannon.  These models fail to address how the gas becomes pressurized prior to firing the cannon.  Our implementation of a compressed air cannon required the use of a valve to create a reservoir of high-pressure gas that is then released to accelerate the projectile.  We propose a new model that takes into account the air flow through the valve.  Our data is in better agreement with this new model than with the prior proposals.
\begin{figure}[htbp] %  figure placement: here, top, bottom, or page
   \centering
   \includegraphics[width=9cm,keepaspectratio]{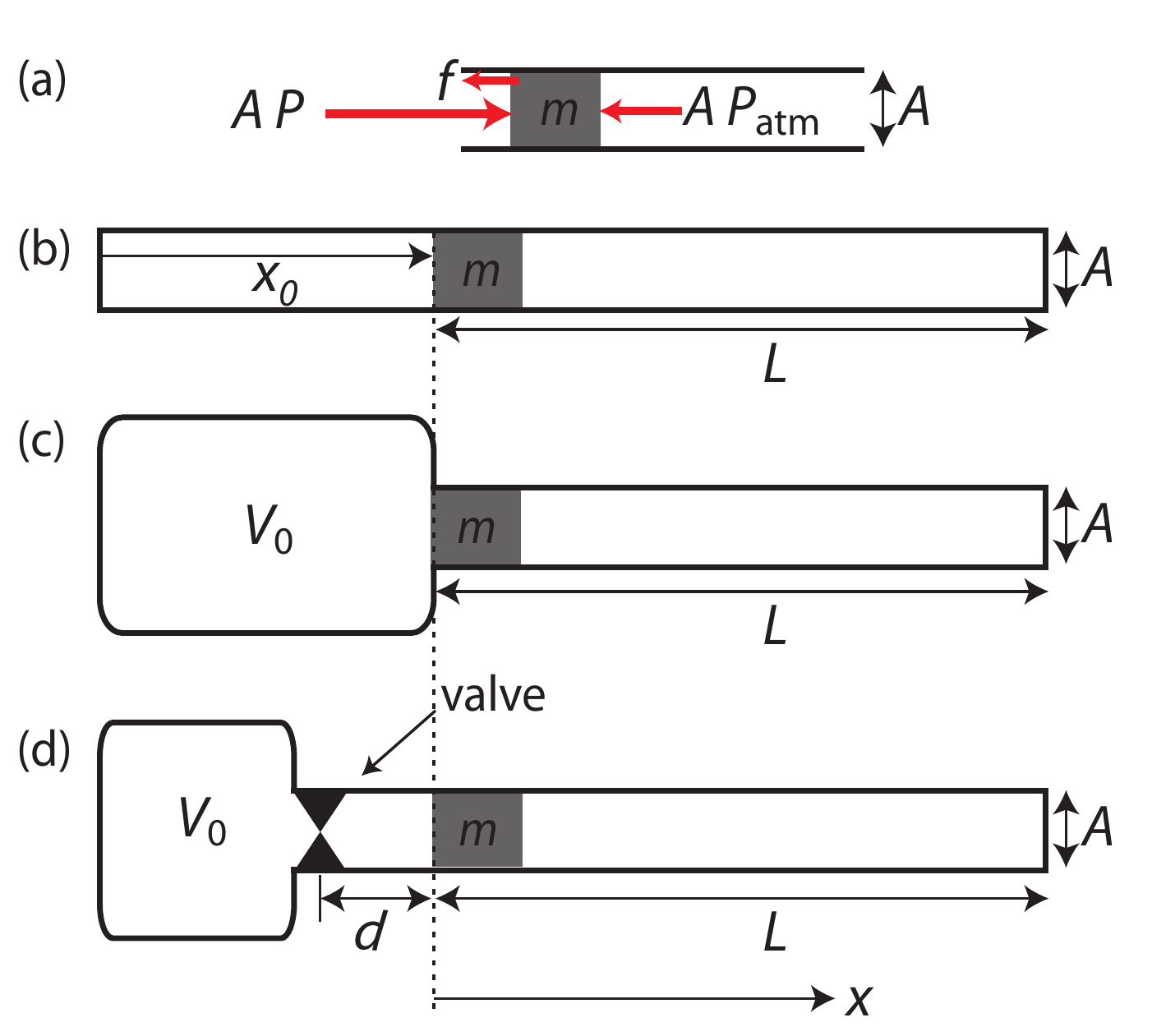} 
   \caption{ (a) The forces on the slug in the barrel of a compressed air cannon.  The pressure $P$ is due to the compressed gas, which expands according to one of these three models: (b) An explosive gas expands adiabatically.\cite{m}  This model is described by Eq.~(\ref{adiabatic}).  (c) Gas from a pressurized reservoir expands isothermally.\cite{d}  See Eq.~(\ref{isothermal}).  (d) The expansion of the gas is limited by a valve with a finite flow factor.  This is the model advocated by this paper.}
   \label{fig:models}
\end{figure}

We begin by reviewing the adiabatic and isothermal gas expansion models.  We want to know the the exit speed $v$ of a projectile of mass $m$, launched from a cannon with initial gas pressure $P_0$.  We model the cannon as a reservoir of volume $V_0$ connected to a long barrel of cross-sectional area $A$ loaded with the projectile (see Fig.~\ref{fig:models}).  As the pressurized gas in the reservoir expands, the gas provides a force to propel the projectile along the length $L$ of the barrel before it exits the barrel.  The total force on the projectile in the barrel is modeled as the sum of the force from the gas in the reservoir $A P(x)$, the force from the air in the barrel at atmospheric pressure $A P_\text{atm}$, and a small linear frictional force $f$ between the projectile and the wall of the barrel, as illustrated in Fig.~\ref{fig:models}(a).  The equation of motion is thus
\begin{align}
\label{newton2} F = m \frac{d^2x}{dt^2} = mv \frac{dv}{dx} = AP(x) - A P_\text{atm} - f \, .
\end{align}
In both gas expansion models, the cannon reservoir volume increases as the projectile moves down the barrel: $V(x) = (V_0 + A x)$.  If we assume an adiabatic expansion as illustrated in Fig.~\ref{fig:models}(b), we know that $P(x) (V_0 + A x)^\gamma = P_0 V_0^\gamma$, where $\gamma = 7/5$ for diatomic gases such as air and $V_0 = A x_0$.  Thus, in the adiabatic case, we have
\begin{align}
mv_\text{ad} \frac{dv_\text{ad}}{dx} = A \left( \frac{P_0 V_0^{\gamma}}{(V_0 + Ax)^{\gamma}} -P_\text{atm} \right) - f \, ,
\end{align}
which yields an exit velocity at $x=L$ of
\begin{align}
\label{adiabatic} v_\text{ad}=\sqrt{\frac{2}{m} \left( \frac{P_0 V_0}{\gamma - 1} \left(1- \left(\frac{V_0}{A L + V_0}\right)^{\gamma - 1}\right)- AL P_\text{atm}- Lf \right)} \, .
\end{align}

The second model, illustrated in Fig.~\ref{fig:models}(c), models the expansion of the gas to be quasistatic and isothermal.  Since $P(x) (V_0+A x) = P_0 V_0$,  the exit velocity is
\begin{equation}
\label{isothermal} v_\text{is} =\sqrt{\frac{2}{m} \left( P_0 V_0 \ln\left(1+\frac{AL}{V_0}\right) -AL P_\text{atm}-  L f \right)} \,  .
\end{equation}

We find, however, that both of these models are over-simplified descriptions of real cannons.  A real pneumatic air cannon has a valve between the reservoir and the barrel in order to allow pressurization of the reservoir before firing the projectile.  While one could imagine a perfect valve that does not have any appreciable effect on the air that flows past it, this is hard to realize in practice.  Since the air flow through a real valve is a function of the pressure drop across the valve, it is unreasonable to ignore the effect of the valve: the pressure in the barrel is not necessarily the same as the pressure in the reservoir.

We propose a new model, shown in Fig.~\ref{fig:models}(d), that takes into account the flow rate of air through the valve.  According to ref.~[\cite{v}],the molecular flow rate $Q$ through the valve is a function of the ratio $\xi$ between the reservoir pressure $P(t)$ and the pressure in the barrel $P_b(t)$, defined as
\begin{equation}
\xi\equiv \frac{P(t)-P_b(t)}{P(t)}\, .
\end{equation}
When the pressure difference is large enough, the pressure ratio saturates to a value of $\xi\rightarrow \xi_\text{max}$, limited by the geometry of the valve and typically between 0.2 and 0.9.  There are thus two flow regimes.  In the non-choked regime ($P(t) < P_b (t)/(1-\xi_\mathrm{max})$), flow is modeled as a function of the pressure differential between the tank pressure and the barrel pressure:
\begin{equation}
Q = \tilde{N} P(t) C_v \left( 1 - \frac{\xi}{3\xi_\mathrm{max}} \right)\sqrt{\frac{\xi}{G_g T Z}}\, .
\label{lowflow}
\end{equation}
In the choked regime  ($P(t) \geq P_b (t)/(1-\xi_\mathrm{max})$), the flow is limited by the geometry of the valve:
\begin{equation}
Q = \tilde{N} P(t) C_v \left( \frac{2}{3} \right)\sqrt{\frac{\xi_\mathrm{max}}{G_g T Z}}\, .
\label{highflow}
\end{equation}
In both of these equations, $G_g=1$ is the specific gravity of air, $T\approx 293$ K is the temperature in the reservoir which is assumed to be constant, and $Z\approx1$ is the compressibility factor.  The flow coefficient $C_v$ of the valve is a unitless parameter that describes the flow capacity of the valve.  Finally, $\tilde{N}=3.11\times10^{19}$~molecules$\cdot\sqrt{\mathrm{K}}/$(Pa$\cdot$s) is an engineering parameter that converts the pressure into flow rate units.\cite{v}  

After having taken into account the valve, we model the gas expansion in the barrel and the tank using the Ideal Gas Law:
\begin{align}
P(t) V_0 &= N(t) k_B T \, , \\
P_b(t) A \left(d+ x(t)\right) &= N_b(t) k_B T \, ,
\end{align}
where $N$ ($N_b$) is the number of molecules in the tank (barrel).  The number of molecules in the tank and barrel are governed by the flow of molecules between them through the valve:
\begin{align}
\frac{dN}{dt} &= -Q \, , \\
\frac{dN_b}{dt} &= Q \, .
\end{align}
Also, as above, Eq.~(\ref{newton2}) governs the position of the slug.  These differential equations, when combined with the initial conditions, are numerically solved to give
\begin{equation}
v_\text{valve}=\left.(dx/dt)\right|_{x=L}
\end{equation}
as a function of initial pressure $P_0$.

\begin{figure*}[htbp] %  figure placement: here, top, bottom, or page
   \centering
   \includegraphics[width=16cm,keepaspectratio]{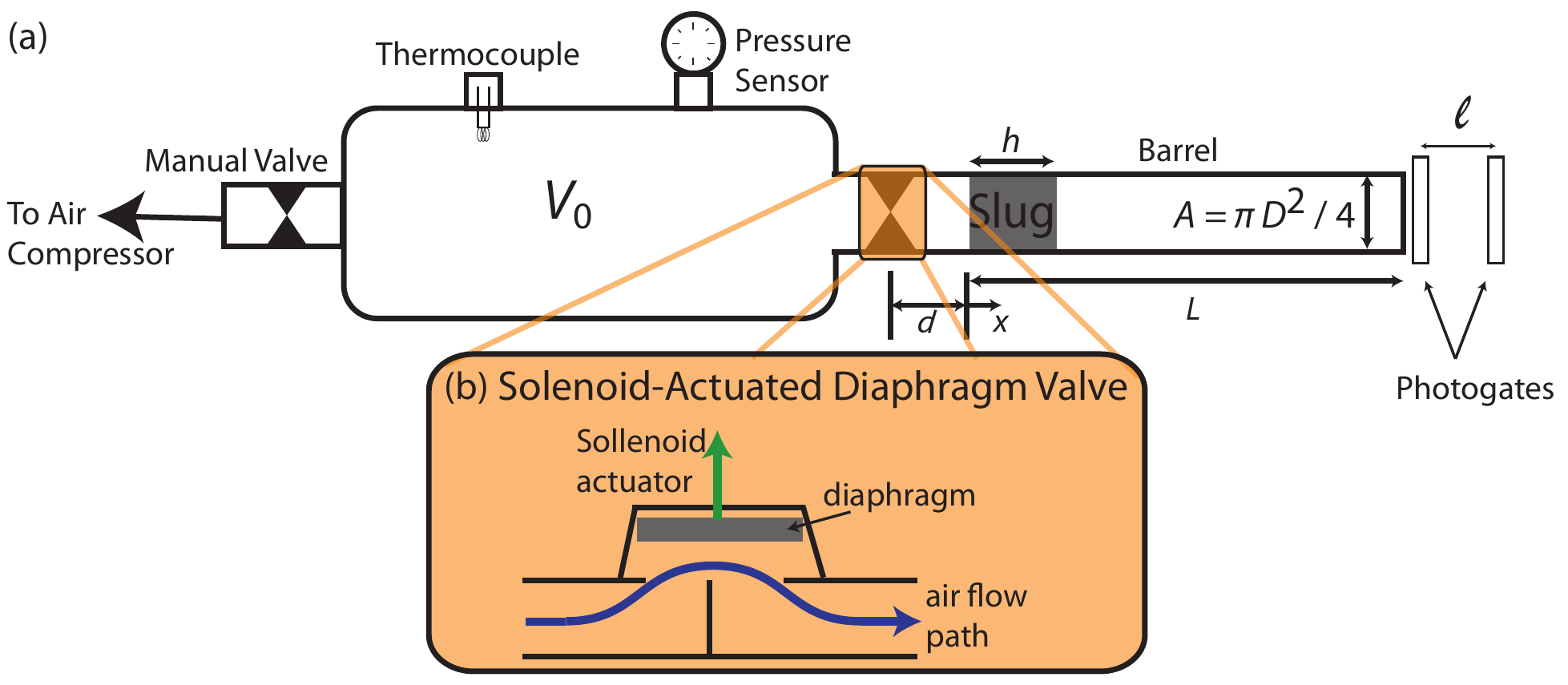} 
   \caption{(a) A schematic of the air cannon. The tank is a reservoir with a volume $V_0=4.196 \pm .010$~L, initially charged to pressure $P_0$.  (b) The tank is discharged using a diaphragm valve with an associated flow factor $C_v$. The pressurized air then propels the slug, with a height $h=4.8019\pm .0006$~cm and mass $m=19.40 \pm .05$~g, a distance of $L=88.25\pm.11$~cm out of the barrel of cross sectional area of $A=2.87233 \pm .00003$~cm$^2$ according to Eq.~(\ref{newton2}).  The exit velocity of the slug is determined using the two photogates at the end of the barrel spaced $\ell= 24.6 \pm 0.9$~cm apart.}
   \label{fig:setup}
\end{figure*}

To test this model, we constructed an air cannon using a steel air tank with a volume of $V_0=4.196 \pm .010$~L (all measurements given to a 95\% confidence interval) as the pressure reservoir. We attached a silicon cell pressure transducer (Omegadyne Model PX309-100GV), a thermocouple (Omega Model TC-K-NPT-E-72), a solenoid-actuated diaphragm valve (Granzow Model 21HN5KY160-14W), and an air intake hand valve to the tank as shown in Fig.~\ref{fig:setup}.  We used a seamless stainless steel (304/304L) threaded pipe for our barrel with a diameter of $1.913 \pm .013$~cm and a total length of $91.6 \pm 0.2$~cm.  We measured the exit velocity using two optical photogates: one positioned at the end of the barrel, the other $\ell=24.6 \pm 0.9$~cm away from the first.  The diaphragm valve opens when a current of $440$~mA  activates a solenoid in the valve.  Data acquisition was triggered when an ammeter connected to the solenoid actuator circuit read an increasing current across the $5$ mA level.

We loaded the cannon with a low-friction cylindrical plastic (acetal copolymer) projectile of mass $m=19.40 \pm .05$~g, height $h=4.8019\pm .0006$~cm, and diameter $D=1.9124 \pm .0006$~cm.  The diameter of the projectile was such that it just fit into the barrel of the cannon.  We tested whether air could escape from around the edges of the projectile by closing the diaphragm valve and attempting to load the cannon.  The projectile was sufficiently airtight that it built back pressure when we tried to insert it.

We loaded the cannon by sliding the projectile into the barrel using a steel rod to push it in to a specific length $L=88.25\pm.11$~cm.  We used an air compressor to pressurize the tank to the desired initial pressure $P_0$.  We typically waited two to three minutes after closing the hand valve to allow the pressure reading in the reservoir to stabilize before opening the diaphragm valve to fire the cannon.

\begin{figure}[htbp] %  figure placement: here, top, bottom, or page
   \centering
   \includegraphics[width=8cm]{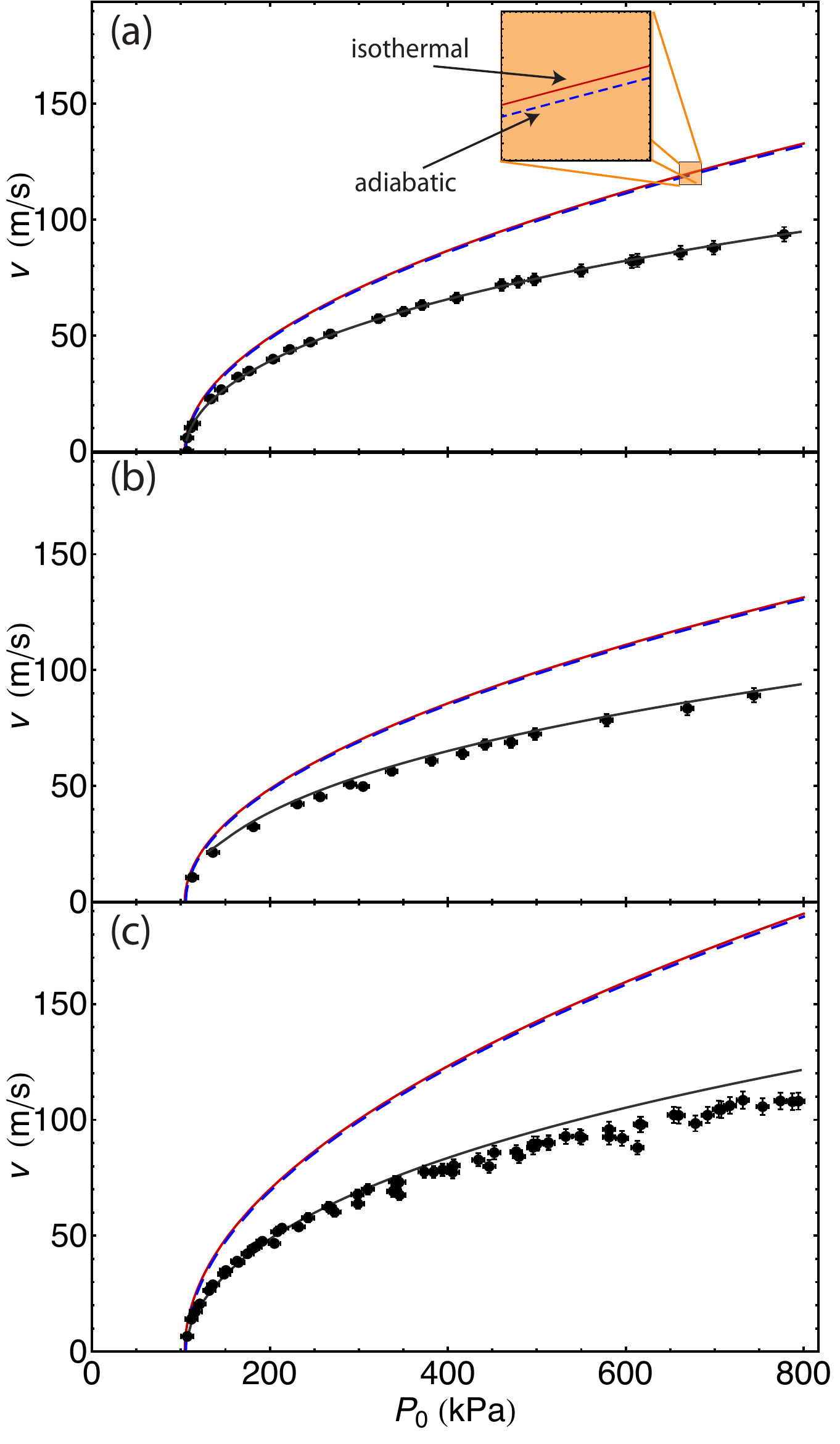} 
   \caption{ 
    (color online.) Exit velocity as a function of initial reservoir pressure of (a) a plastic projectile with $m=19.40 \pm .05$~g and $h=4.8019\pm .0006$~cm,  (b) an aluminum projectile with $m=19.90 \pm .05$~g and $h=2.5923\pm .0006$~cm, and (c) a plastic projectile with $m=9.60 \pm .05$~g and $h=2.3929\pm .0006$~cm. The solid red curve is the isothermal model and the dashed blue curve is the adiabatic model.  The two models are close to each other because the temperature drop associated with the adiabatic expansion is so small.  Our data disagree with both.  The black curve is a numerical plot of our new model fit to the data presented in (a) with parameters $\xi_\text{max}=.80\pm.11$ and $C_v=1.93\pm.04$.}
   \label{fig:originalprojectile}
\end{figure}

We collected data for the exit velocity $v$ of the projectile as a function of initial reservoir pressure $P_0$, shown in Fig.~\ref{fig:originalprojectile}(a).  It is clear that our data are in gross disagreement with both the adiabatic and isothermal models.  However, when we perform a manual two-parameter fit of our data to the valve flow model presented above, we get much better agreement.  Our fit yields the two parameters (with estimated uncertainties) $\xi_\text{max}=.80\pm.11$ and $C_v=1.93\pm.04$.

In all three models, we have ignored the frictional term $f$.  This is because we find that the introduction of a non-zero frictional term in the adiabatic and isothermal models corresponds to a horizontal shift of the model.  Because our data have a very small horizontal offset, we take $f\approx 0$ for all of our calculations.  We also find experimentally that a very small initial tank pressure above atmospheric pressure ejects the projectile, suggesting that $f$ is, indeed, small.

We also performed our experiment with an aluminum projectile of about the same mass and half the length (specifically, $m=19.90 \pm .05$~g and $h=2.5923\pm .0006$~cm) with data shown in  Fig.~\ref{fig:originalprojectile}(b).  The model, using the same fit parameters as above, is in good agreement with the data.

Finally, we performed our experiment with another plastic projectile of approximately half the length and mass of the first (specifically, $m=9.60 \pm .05$~g and $h=2.3929 \pm .0006$~cm.  These data are shown in Fig.~\ref{fig:originalprojectile}(c).  It is evident here that our model overshoots the data for high $P_0$, using the same fit parameters for the model that we found found above.  We believe that this is due to quadratic air drag, which has a larger effect at higher velocities (which occur at higher initial pressures $P_0$).  Our model would need to be improved in order to incorporate this effect.  The reason we would not have seen this issue in Fig.~\ref{fig:originalprojectile}(a-b) is that the larger mass of the slug would mean that the acceleration due to drag would be smaller.

In conclusion, we have shown that both the adiabatic and isothermal expansion models are not consistent with our data.  A gas flow model that accounts for the valve is in much better agreement with the data.  This is because the assumption of both the adiabatic and isothermal models that the air pressure in the reservoir is the same as the pressure in the barrel is difficult to meet given the necessity of using a valve to pressurize the reservoir.  We now have a way to predict exit velocity as a function of initial tank pressure, which will be useful in the development of future undergraduate labs utilizing compressed air cannons.

\acknowledgements{
We would like to thank Daniel Brown, Sam Krutz, and Micah Milliman for their foundational work on designing the air cannon used in this experiment, and Diego Aliaga, Chris Beard, Jacob Castilow, Jon Barlow, and Scott Pond for their preliminary data gathering and analysis.  We would also like to think Dr. James Brown for his help in developing the theory.
}

\end{document}